\documentclass[aps,prl,preprint,superscriptaddress]{revtex4}

\usepackage{graphicx,amsmath}
\usepackage{color}

\usepackage{booktabs}
\usepackage[flushleft]{threeparttable}

\usepackage{amssymb}
\usepackage{amsmath}
\usepackage{float}
\usepackage{psfrag}
\usepackage{siunitx}
\usepackage{ulem}
\usepackage{romannum}

\def\be{\begin{equation}}
\def\ee{\end{equation}}
\def\ba{\begin{eqnarray}}
\def\ea{\end{eqnarray}}

\begin{document}

\title{Evaporation-triggered segregation of sessile binary droplets}
\author{Yaxing Li}
\affiliation{Physics of Fluids group, Department of Science and Technology, Mesa+ Institute, and 
J. M. Burgers Centre for Fluid Dynamics, University of Twente, P.O. Box 217, 7500 AE Enschede, The Netherlands}
\author{Pengyu Lv}
\affiliation{Physics of Fluids group, Department of Science and Technology, Mesa+ Institute, and 
J. M. Burgers Centre for Fluid Dynamics, University of Twente, P.O. Box 217, 7500 AE Enschede, The Netherlands}
\author{Christian Diddens}
\affiliation{Physics of Fluids group, Department of Science and Technology, Mesa+ Institute, and 
J. M. Burgers Centre for Fluid Dynamics, University of Twente, P.O. Box 217, 7500 AE Enschede, The Netherlands}
\affiliation{Department of Mechanical Engineering, Eindhoven University of Technology, P.O. Box 513, 5600 MB Eindhoven, The Netherlands}
\author{Huanshu Tan}
\affiliation{Physics of Fluids group, Department of Science and Technology, Mesa+ Institute, and 
J. M. Burgers Centre for Fluid Dynamics, University of Twente, P.O. Box 217, 7500 AE Enschede, The Netherlands}
\author{Herman Wijshoff}
\affiliation{Department of Mechanical Engineering, Eindhoven University of Technology, P.O. Box 513, 5600 MB Eindhoven, The Netherlands}
\affiliation{Oc\'{e} Technologies B.V., P.O. Box 101, 5900 MA Venlo, The Netherlands}
\author{Michel Versluis}
\affiliation{Physics of Fluids group, Department of Science and Technology, Mesa+ Institute, and  J. M. Burgers Centre for Fluid Dynamics, University of Twente, P.O. Box 217, 7500 AE Enschede, The Netherlands}
\author{Detlef Lohse}
\email{d.lohse@utwente.nl}
\affiliation{Physics of Fluids group, Department of Science and Technology, Mesa+ Institute, and 
J. M. Burgers Centre for Fluid Dynamics, University of Twente, P.O. Box 217, 7500 AE Enschede, The Netherlands}
\affiliation{Max Planck Institute for Dynamics and Self-Organization, 37077 G\"ottingen, Germany}

\begin{abstract} 

Droplet evaporation of multicomponent droplets is essential for various physiochemical applications, e.g. in inkjet printing, spray cooling and microfabrication. In this work, we observe and study phase segregation of an evaporating sessile binary droplet, consisting of a mixture of water and a surfactant-like liquid (1,2-hexanediol).  The phase segregation (i.e., demixing) leads to a reduced water evaporation rate of the droplet and eventually the evaporation process ceases due to shielding of the water by the non-volatile 1,2-hexanediol. Visualizations of the flow field by particle image velocimetry and numerical simulations reveal that the timescale of water evaporation at the droplet rim is faster than that of the Marangoni flow, which originates from the surface tension difference between water and 1,2-hexanediol, eventually leading to segregation. 

\end{abstract}


\maketitle


The evaporation of a sessile droplet has attracted a lot of attention over the past years~\cite{picknett1977,deegan1997,lohse2015rmp,hu2002,popov2005,cazabat2010,sbonn2006,ristenpart2007,lim2008,schoenfeld2008,gelderblom2011,marin2011,brutin2011,ledesma2014,Tan2016}, not only from a fundamental scientific perspective, but also because of many technological and biological applications, such as inkjet printing~\cite{park2006control}, nanopatterning depositions~\cite{kuang2014controllable}, and DNA stretching~\cite{jing1998automated}. Within the whole class of problems, the so-called "coffee-stain effect" which was presented to the scientific community 20~y ago~\cite{deegan1997}, has become paradigmatic. The problem and its variations keep inspiring the community. This holds not only for the evaporation of liquids with dispersed particles~\cite{marin2011,nguyen2017manipulating}, but also for that of liquid mixtures, including binary and ternary mixtures~\cite{Sefiane2003,sefiane2008,kim2016controlled,Tan2016,diddens2017evaporating}. In recent work on an evaporating Ouzo drop (a ternary mixture of water, ethanol and anise oil), Tan $et\ al$.~\cite{Tan2016} showed that a phase transition and the nucleation of oil microdroplets can be triggered by evaporation. The reason for the nucleation lies in the varying solubility of oil in the ethanol-water mixtures: the high evaporation rate at the rim of the droplet together with the higher volatility of ethanol as compared to water causes an oil oversaturation at the rim, leading to localized oil microdroplet nucleation. The oil microdroplets are advected over the whole drop by Marangoni flow and further droplets later nucleate in the bulk. Finally, the microdroplets are jammed and coalesce during the further evaporation process, eventually leading to the formation of a separated oil phase in the remaining binary water/oil droplet. 
Liquid-liquid phase separation during evaporation not only occurs for Ouzo droplets, but is omnipresent in nature and technology~\cite{lobl1994nucleation,wang1998nucleation,rao1989nucleation,hyman2014liquid}. 

In this work, we study segregation within an evaporating 1,2-hexanediol/water miscible binary droplet. 1,2-hexanediol is used in a variety of applications, such as co-surfactant for modifying the sodium dodecyl sulfate (SDS) micelles~\cite{kennedy2001interaction} and oil solubilization in ternary systems~\cite{d2003small}. The features of its aqueous solution are widely studied in many previous papers~\cite{hajji1989comparative,frindi1991ultrasonic,Szekely2007}, which show that 1,2-hexanediol molecules form micelle-like aggregates characterized by a critical micelle concentration (CMC) in aqueous solutions, leading to an almost constant surface tension above the CMC~\cite{Romero200767}. Compared with water, 1,2-hexanediol is non-volatile under room conditions, implying a preferred evaporation of the more volatile water during the drying process. However, to the best of our knowledge, the segregation of the miscible 1,2-hexanediol and water during the evaporation process has never been observed, nor studied. In this paper, we explore experimentally and numerically the mechanism of segregation of 1,2-hexanediol from the miscible water, that is found to be triggered by selective evaporation. 

\begin{figure}[h]
\centering
\includegraphics[width=1\textwidth]{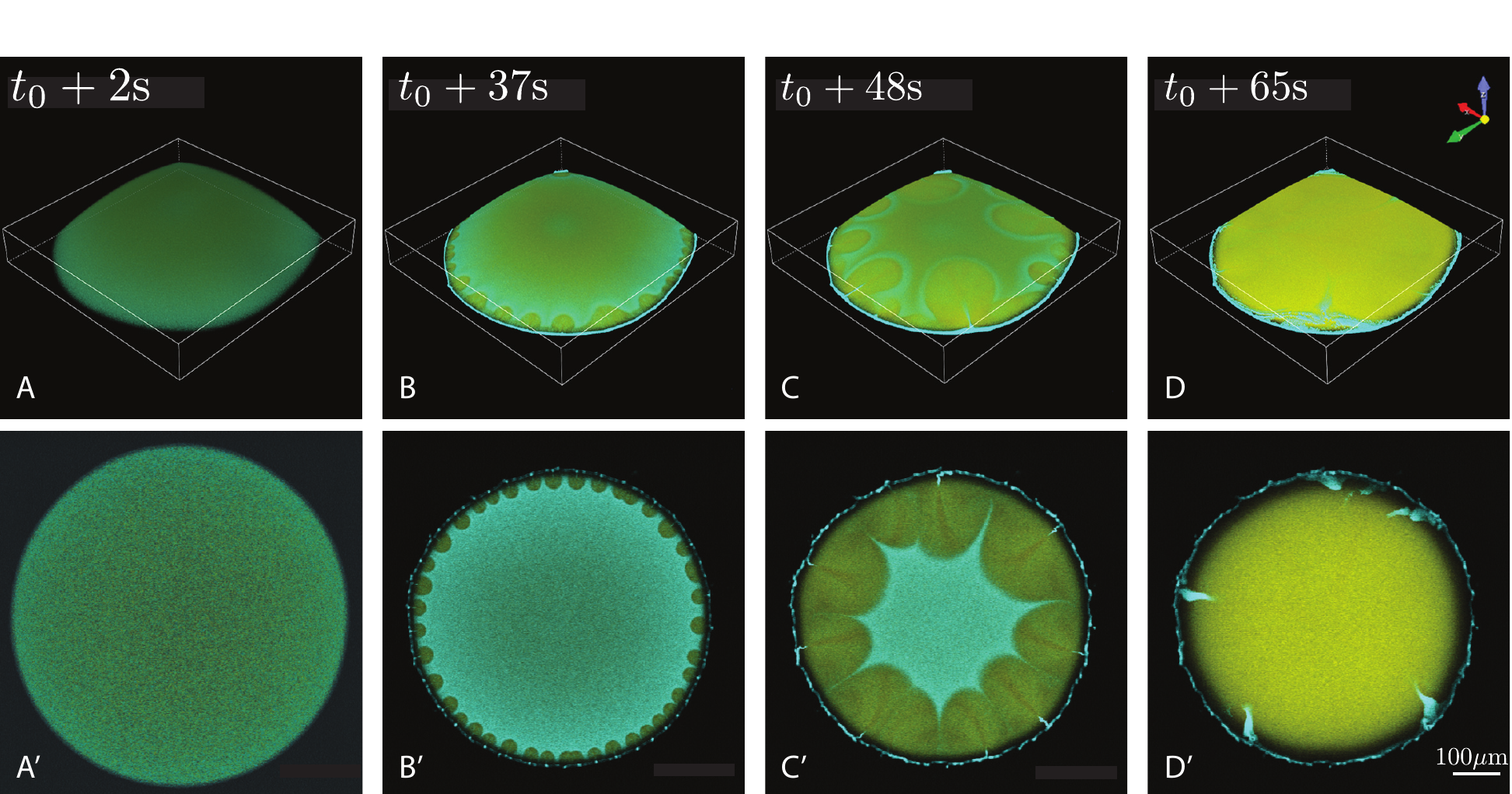}
\caption{Confocal images of the segregation process during droplet evaporation in a side (A-D) and bottom (A'-D') view taken at the same times. (A-D) The confocal microscope scans the rectangular box with the volume 590 $\mu \text{m}$ $\times$ 590 $\mu \text{m}$ $\times$ 90 $\mu \text{m}$. Water (blue) and 1,2-hexanediol (yellow) are labeled with different dyes for the observation. (A and A') In the beginning, the droplet is homogeneously mixed. (B and B') At about 34~s after recording started, 1,2-hexanediol nucleates at the contact line of the droplet, which is revealed as yellow round shapes. (C and C') The nucleated microdroplets of 1,2-hexanediol gradually grow and coalesce. (D and D') The evaporation ends when 1,2-hexanediol fully covers the surface of the droplet.
 }
\label{fig:images_conf}
\end{figure}

We begin with the visualization of the distribution of the mixture components during evaporation by labelling water and 1,2-hexanediol with the fluorescent dyes dextran and nile red, respectively. A dyed 0.5 $\mu\text{L}$ binary droplet with initial 10$\%$ mass concentration of 1,2-hexanediol (around the CMC~\cite{Romero200767}) is deposited on a transparent hydrophobic octadecyltrichlorosilane (OTS)-glass surface, while monitoring its evaporation under ambient conditions with confocal microscopy from side and bottom (see supporting information). The contact angle of the droplet varies between 43$^\circ$ and 23$^\circ$ during the whole evaporation process, measured by bright-field imaging from side view. Fig.\  \ref{fig:images_conf} presents the segregation process of the evaporating binary droplet. In the beginning the droplet is homogeneously mixed, as revealed by the uniformed green colour over the surface and on the bottom (Fig.\  \ref{fig:images_conf}$A$,\ref{fig:images_conf}$A'$).  About 34~s after deposition, 1,2-hexanediol microdroplets nucleate at the rim of the droplet, revealed by the yellow colour (Fig.\  \ref{fig:images_conf}$B$,\ref{fig:images_conf}$B'$). During further evaporation, the nucleated microdroplets of 1,2-hexanediol grow and coalesce, which forms star-shape binary mixture area revealed in blue colour (Fig.\  \ref{fig:images_conf}$C$,\ref{fig:images_conf}$C'$). Eventually, 1,2-hexanediol covers the whole surface of the droplet and the evaporation process stops with some water being entrapped by the 1,2-hexanediol (Fig.\  \ref{fig:images_conf}$D$,\ref{fig:images_conf}$D'$). From comparing the initial and the final size, we calculate that approximately 96$\%$ of the water has evaporated while 4$\%$ got trapped. 

To obtain insight into the segregation process, we record the evolution of the flow field within the evaporating binary 1,2-hexanediol/water droplet by particle image velocimetry (PIV) combined with confocal microscopy. For a first qualitative understanding, we added 1 $\mu m$ diameter fluorescent particles at a concentration of $5 \times 10^{-5} \ \text{vol}\%$, which is much less than the particle concentration required for a quantitative PIV measurement~\cite{diddens2017evaporating,kim2016controlled}. The whole droplet and all particles were illuminated: particles near the substrate (pink colour) were in focus of the camera; the grey or transparent objects were out-of-focus particles and reside in the upper part of the droplet.
 
\begin{figure}[h]
\centering
\includegraphics[width=1\textwidth]{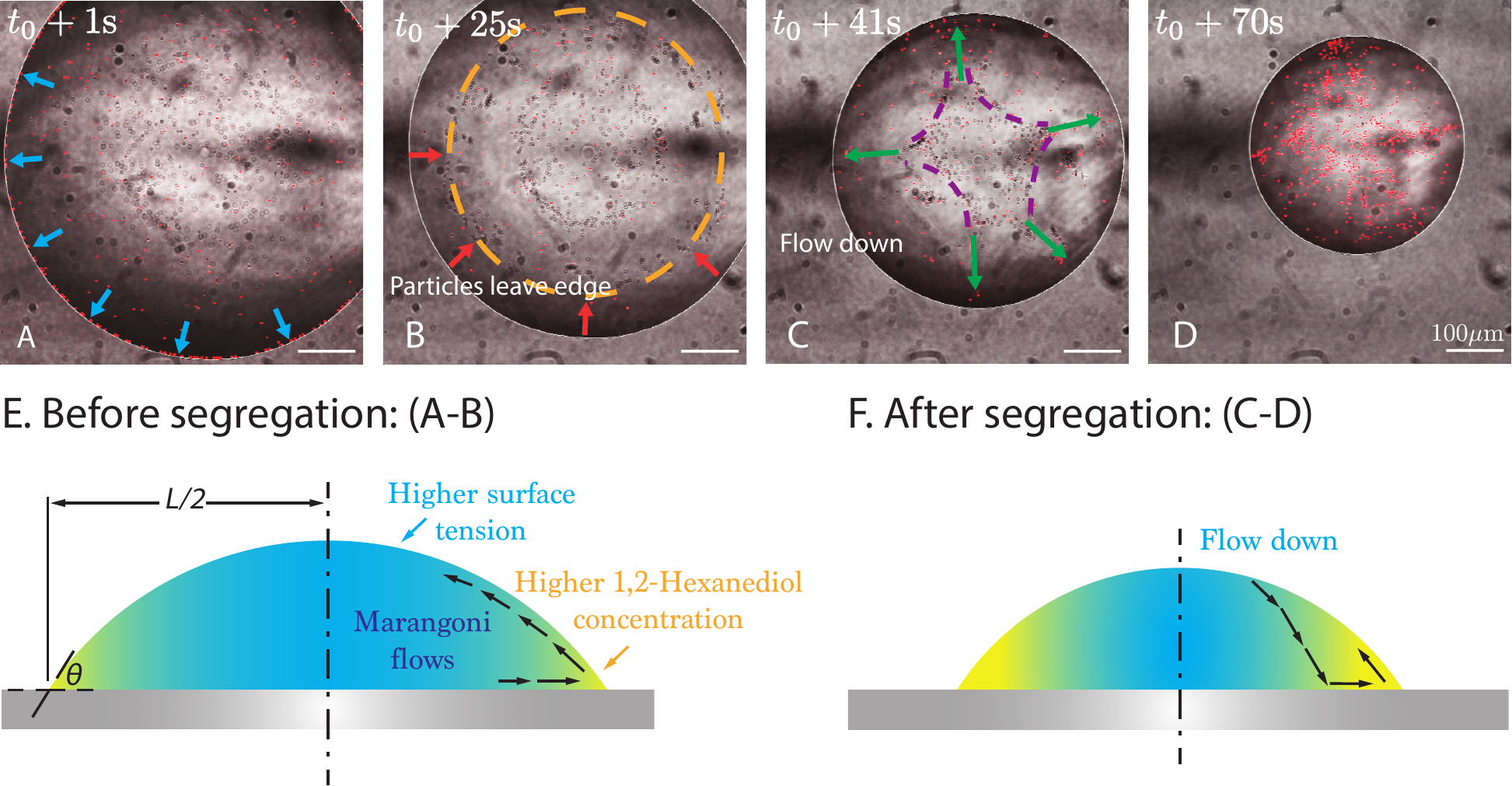}
\caption{(A-D) Bottom-view snapshots of the droplet seeded with fluorescent particles in different life phases. (A) The flow is directed radially outwards near the substrate, as shown by particles transported to the contact line (blue arrows). (B) All the particles are released from the contact line and flow to the upper center (orange circle) of the droplet (red arrows). (C) Particles floating on the upper layer form a star shaped pattern (purple lines) and flow downward through the fingers of the star (green arrows). (D) When the droplet stops evaporating, the particles are deposited homogeneously on the substrate, without leaving a coffee-stain. (E-F) Schematics of the flow inside the binary droplet at different phases. (E) Before segregation, the surface tension gradient drives a Marangoni flow from the edge to the apex of the droplet. (F) After segregation, the nucleated microdroplets of 1,2-hexanediol grow and coalesce. At the same time, water-rich liquid from the upper layer of the droplet flows down through the streams between neighbouring nucleated microdroplets.
 }
\label{fig:PIV_schematics}
\end{figure}

Initially, the flow is directed radially outwards near the substrate (see Fig~\ref{fig:PIV_schematics}A). In this phase, only water evaporates from the binary droplet and the droplet is thin, $H/L \ll 1$, where the droplet height $H$ is approximately 60 $\mu$m and droplet footprint diameter $L$ is about 600  $\mu$m. Therefore, due to the relative high concentration of 1,2-hexanediol caused by the singularity of the water evaporation rate at the rim of the sessile droplet~\cite{cazabat2010}, a Marangoni flow is driven from the contact line to the apex of the droplet by the surface tension gradient, which originates from the concentration variation along the surface. Note that the surface tension of 1,2-hexanediol aqueous solution is monotonously decreasing with 1,2-hexanediol concentration when it is lower than the CMC~\cite{Romero200767}. As a consequence, a convective flow inside the droplet is driven by the Marangoni flow and water is transported to the contact line by radial outflow near the substrate. However, here the convective flow within the droplet is not sufficient to compensate for the evaporative water loss near the contact line. The typical outwards flow velocity shortly after deposition is $U \approx 1\ \mu$m/s, implying a Reynolds number $Re = \rho HU/\mu \approx 10^{-5}$, where $\rho \approx 10^3$ kg/m$^3$ is the liquid density and $\mu \approx 10$ mPa~s is the viscosity. We compare the time scales of evaporation $t_{\text{ev}} \sim \rho LH/(D_{\text{w},\text{air}} \Delta c_w) $~\cite{dietrich2016} with that of convective Marangoni flow $t_{\text{Ma}} \sim L/U $ on the surface: $t_{\text{ev}}/t_{\text{Ma}} \sim \rho HU/(D_{\text{w},\text{air}} \Delta c_\text{w}) \approx 10^{-1}$, where $D_{\text{w},\text{air}} = 2.4 \times 10^{-5}$ m$^2$/s is the diffusion coefficient of water vapor at room temperature and $\Delta c_\text{w} \approx 10^{-2}$ kg/m$^3$ is the vapor concentration difference from the air-liquid interface to the surrounding air. The small ratio $t_{\text{ev}}/t_{\text{Ma}} \ll 1$ indicates that the water loss due to the evaporation cannot be replenished by convective flux. Therefore, the concentration of 1,2-hexanediol near the contact line keeps increasing due to the insufficient compensation by the water due to the low convective flow. 

In the second phase, after about 18~s (Fig.~\ref{fig:PIV_schematics}B), all particles, which had accumulated at the contact line, released and simultaneously moved upward along with the Marangoni flow~\cite{marin2016surfactant,kim2016controlled}. They move along the liquid-air interface due to the hydrophilicity of the particles and the diol accumulation at the rim. In the third regime (Fig.~\ref{fig:PIV_schematics}C), the particles floating on the upper layer formed a star shape which is revealed by the orange dashed line, and then flowed down to the bottom of the droplet through the shape of fingers of a star. Compared with the observations in Fig.\ref{fig:images_conf}, the star shape corresponds to the blue part in Fig.\ref{fig:images_conf}C and C', which represents the water-rich area. The fingers are formed by the liquid on the upper layer flowing downward through the streams between each two neighbouring growing nucleated microdroplets. During the segregation process, the surface tension force is dominant compared to gravity forces, as the Bond number Bo = $\rho g L^{2}/\sigma \approx 10^{-1}$, where $g = 9.8\ \text{m}/\text{s}^{2}$ gravity and $\sigma \approx 24$ mN/m is the surface tension of the 1,2-hexanediol aqueous solution above the CMC~\cite{Romero200767}.  In the final phase, when 1,2-hexanediol almost entirely covers the surface and the evaporation ceases, particles flow irregularly and eventually are deposited uniformly with no particles accumulating at the edge when evaporation fully stops (Fig.~\ref{fig:PIV_schematics}D). 

\begin{figure}[h]
\centering
\includegraphics[width=1\textwidth]{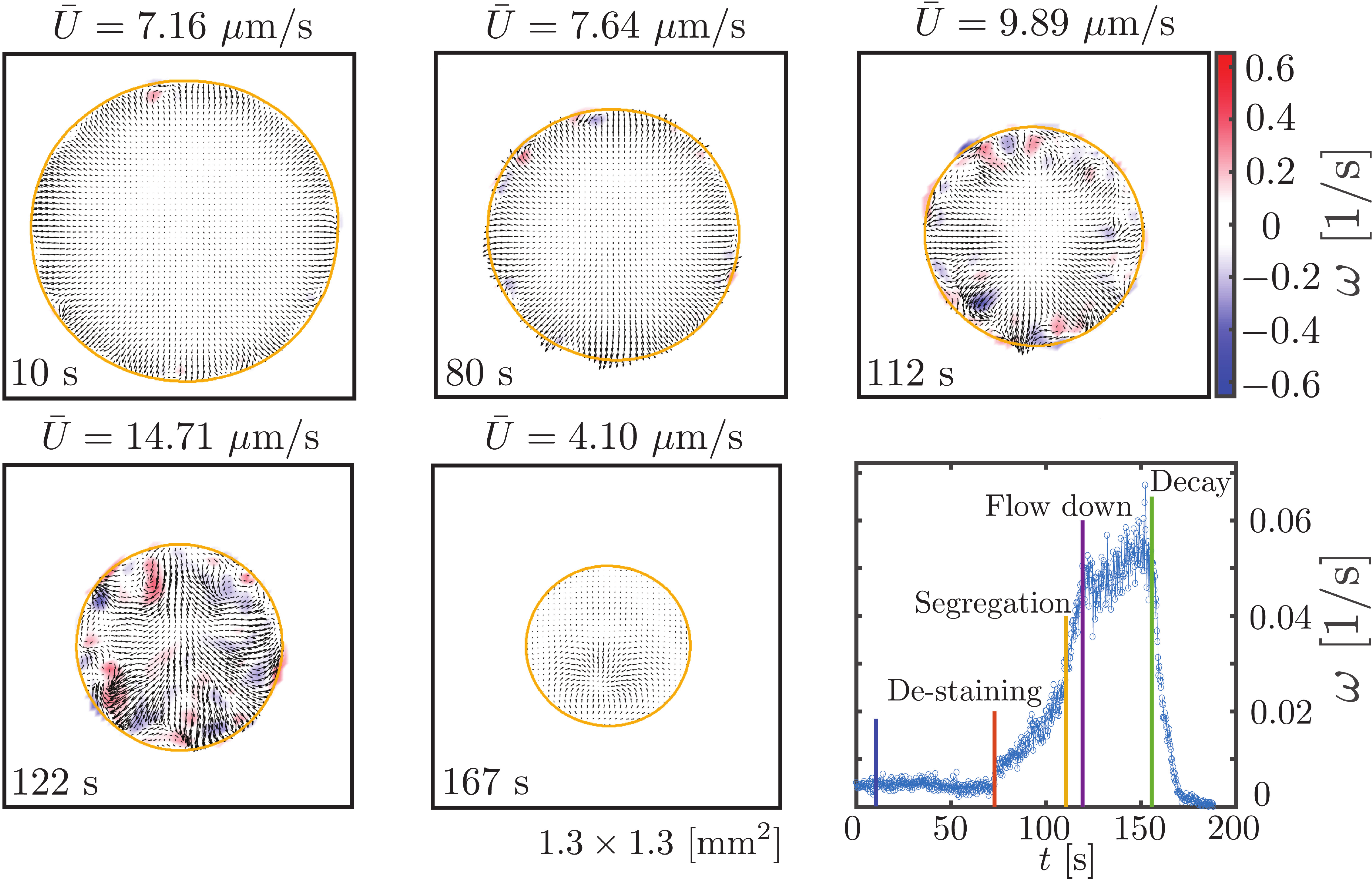}
\caption{Particle image velocimetry results showing the flow field near the substrate in terms of velocity vectors and vorticity, allowing to identify evaporation stages. The five vertical lines show the moments of the five snapshots. (Blue line: 10s; red line: 80s; yellow line: 112s; purple line: 122s; green line: 167s.) 
 }
\label{fig:PIV_plot}
\end{figure}

To obtain a quantitative analysis of the flow field during evaporation, we add 520 nm diameter fluorescent particles at a concentration of $2 \times 10^{-3}\ \text{vol}\%$ into the droplet. The flow speed $U$ and the wall-normal vorticity $\omega = \partial_x u_y - \partial_y u_x$ for the in-plane velocity ($u_x, u_y$) are measured during the whole evaporation process. Also from the evolution of the mean vorticity $\bar{\omega}$, the different life phases of the evaporation can be identified, now even quantitatively, see Fig.~\ref{fig:PIV_plot}. In the early phase, there is almost only outward radial flow, resulting in constant low vorticity. After de-staining of the particles, there are some small vortices appearing near the droplet rim due to the receding contact line. When segregation starts, the vorticity sharply increases due to a series of vortices forming in the nucleated microdroplets of 1,2-hexanediol, see also in Fig.~\ref{fig:images_conf}C. During coalescence of the growing nucleated microdroplets, small vortices merge and form larger vortices. When the growing microdroplets reach the area where floating particle reside, the particles flow down to the bottom. Finally the flow becomes irregular and then vanishes at the end of the evaporation process.

Sessile droplet evaporation is a diffusion-dominated process driven by the concentration gradient of the droplet's constituent(s) from the droplet interface towards the surroundings. The case of a pure evaporating sessile droplet has analytically been solved by Popov~\cite{popov2005}, see the supporting information.

For a droplet consisting of more than one component, the situation gets more complicated and can only be treated numerically. Several generalizations are necessary to adopt Popov's model to a multi-component droplet. Since these generalizations are described in detail in several recent publications \cite{Diddens2017a,Diddens2017b,diddens2017evaporating,Tan2016,Tan2017a}, only a brief overview of the model is given in the following, focusing on the case of the present binary mixture.

As 1,2-hexanediol is non-volatile, only the evaporation rate of water has to be determined. However, in contrast to the case of a pure water droplet, where the water vapor concentration $c_{\text{w}}$ is saturated directly above the liquid-gas interface, i.e. $c_{\text{w}}=c_{\text{w,s}}$,  in the case of a droplet, consisting of two miscible liquids, the vapor concentration is given by the vapor-liquid equilibrium. This equilibrium can be expressed by Raoult's law, i.e. $c_{\text{w}}=\gamma_{\text{w}}X_{\text{w}}c_{\text{w,s}}$, where $X_{\text{w}}$ is the mole fraction of water in the liquid and $\gamma_{\text{w}}$ is the activity coefficient of water for the 1,2-hexanediol/water mixture. The water vapor concentration is in general non-uniform along the interface and changes over time. The evaporation process is modeled by the quasi-steady vapor diffusion equation $\nabla^2c_{\text{w}}=0$. We use Raoult's law at the liquid-gas interface and the ambient vapor concentration $c_{\text{w}}=c_{\text{w},\infty}$ far away from the droplet as Dirichlet boundary conditions. 
The evaporation rate of water $J_{\text{w}}$ is then given by the diffusive flux at the interface, i.e. $J_{\text{w}}=-D_{\text{w,air}}\partial_n c_{\text{w}}$.

In case of a pure droplet, or for a multicomponent droplet in the presence of a very intense Marangoni flow, it is sufficient to keep track of the total mass of each species over time to predict the volume evolution \cite{Diddens2017a,Tan2017a}. Here, however, the Marangoni flow is weak and segregation occurs, so that an explicit spatio-temporal dependence of the local liquid composition emerges. Hence, the convection-diffusion equation for the water mass fraction $Y_{\text{w}}$ has to be solved inside the droplet:
\begin{equation}
\rho\left(\partial_t Y_{\text{w}} + \vec{u}\cdot\nabla Y_{\text{w}}\right) = \nabla\cdot\left(\rho D \nabla Y_{\text{w}} \right) - J_{\text{w}}\delta_{\text{interf.}}
\label{eq:num:convdiffuyw}
\end{equation}
The mass density of the liquid $\rho$ and the diffusivity $D$ are composition-dependent quantities, i.e. $\rho(Y_{\text{w}})$ and $D(Y_{\text{w}})$. The evaporation rate of water enters Eq. \eqref{eq:num:convdiffuyw} as interfacial sink term $\delta_{\text{interf.}}$. 

The advection velocity $\vec{u}$ is obtained from the Stokes equation, subject to a no-slip boundary condition at the substrate, the kinematic boundary condition considering evaporation, the Laplace pressure in normal direction at the liquid-gas interface, and the Marangoni shear stress that arises due to the composition-dependent surface tension $\sigma(Y_{\text{w}})$ in tangential direction at the liquid-gas interface. Furthermore, the composition-dependence of the dynamic viscosity $\mu(Y_{\text{w}})$ has to be considered. For the composition-dependence of the liquid's material properties, we have fitted experimental data and/or used models. More details and plots of these relations can be found in the supplementary information. 

\begin{figure}[h]
\centering
\includegraphics[width=1\textwidth]{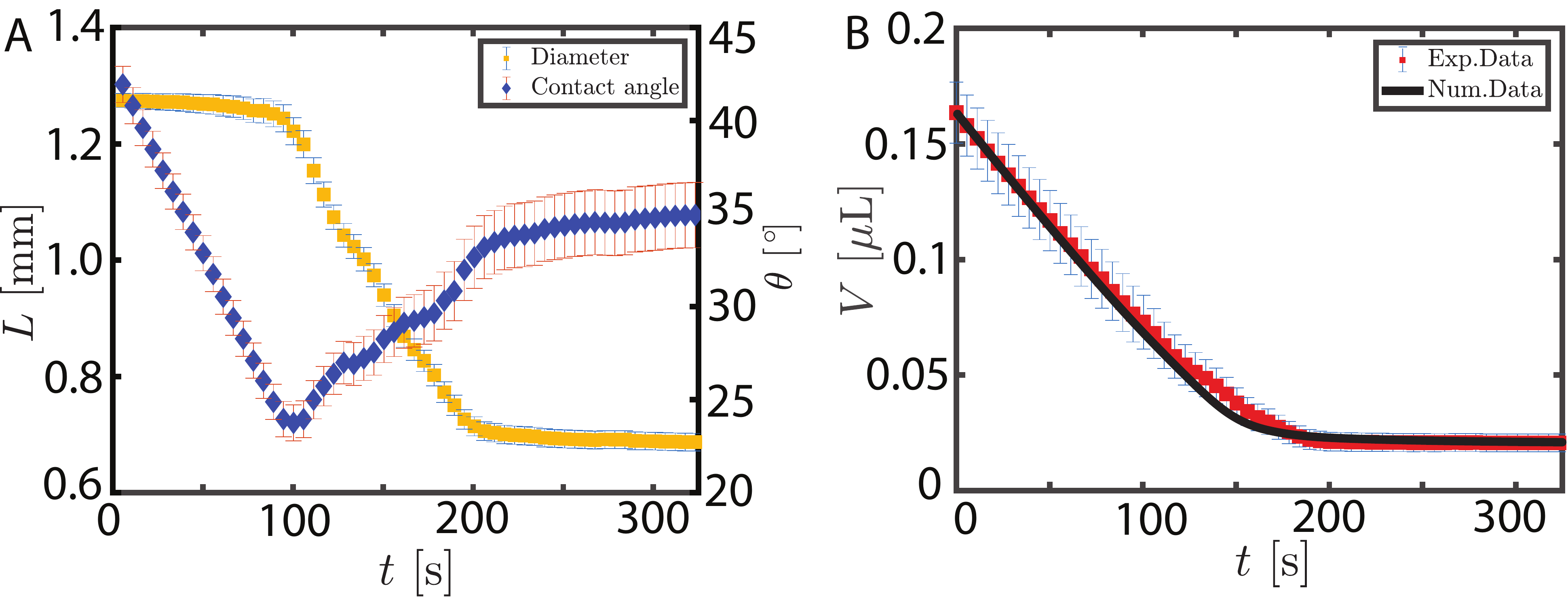}
\caption{Experimental (data points) and numerical results (solid line) for the temporal evolution of the geometrical parameters (A): footprint dimater $L$, contact angle $\theta$ and (B): volume $V$ from experiment and numerical simulation. The error bars are deduced from the experimental accuracy.}
\label{fig:parameters_10}
\end{figure}

The resulting set of coupled equations can be solved numerically with a finite element method \cite{Diddens2017b,diddens2017evaporating,Tan2017a}. We restrict ourselves to axial symmetry. Since the evolution of the contact angle is determined by microscopic interactions at the contact line, it cannot be predicted by the model. Instead, the experimentally measured evolution of the contact angle was imposed throughout the simulation, see Fig.~\ref{fig:parameters_10}A.

\begin{figure}[h]
\centering
\includegraphics[width=1\textwidth]{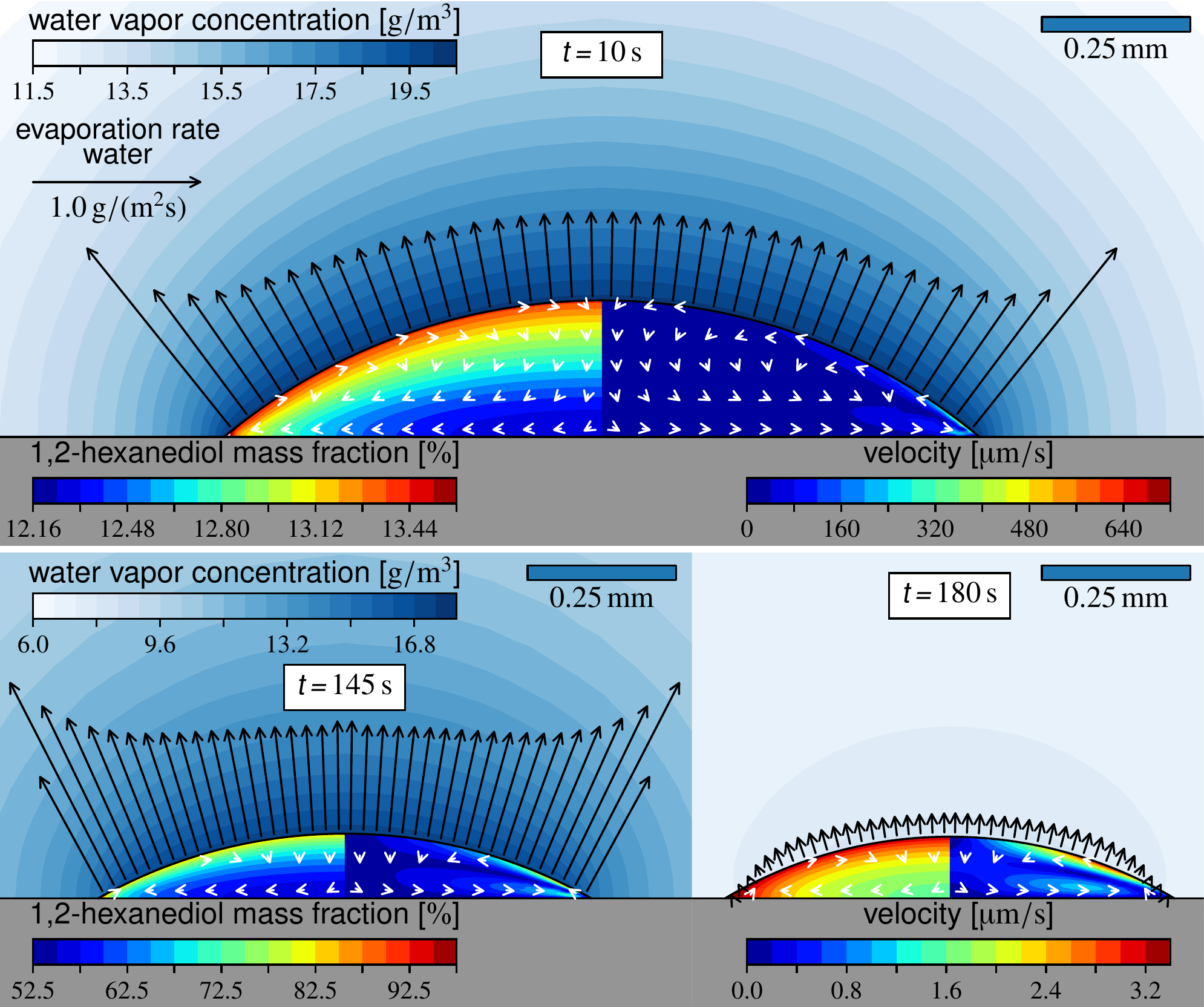}
\caption{These snapshots from the simulation of an evaporating droplet with initially $10 \%$ of 1,2-hexanediol with the axisymmetric finite element model at different times $t$. In the gas phase, the water vapor concentration $c_{\text{w}}$ is shown and the corresponding evaporation rate $J_{\text{w}}$ is indicated by the arrows at the interface. Inside the droplet, the mass fraction of 1,2-hexanediol (left) and the velocity (right) is depicted. Note the very different phenomena at $t = 10$s (upper) and at the two later times $t = 145$s, and $t = 180$s (lower).}
\label{fig:10per}
\end{figure}

In Fig.~\ref{fig:10per}, these snapshots of the simulation for the droplet consisting of an initial $10 \%$ 1,2-hexanediol are depicted. While initially a considerable Marangoni flow is present and the profile of the evaporation rate resembles the case of a pure water droplet, the situation drastically changes at later times: The Marangoni flow ceases due to the nearly constant surface tension at lower water concentrations. Towards the end of the evaporation process, the evaporation rate suddenly decreases once the water concentration $Y_{\text{w}}$ falls below a threshold of about $10 \%$. Since this transition sets in near the contact line, the profile of the evaporation rate shows a remarkable deviation from the case of a pure droplet with a pronounced evaporation at the apex in this stage. The evaporation-triggered segregation effect in radial direction is well captured by the model. Finally, a remaining water residue is entrapped in the bulk of the droplet (4$\%$ of the initial water content) which can only reach the interface by diffusion. The comparison between simulation and experimental data shows an excellent quantitative agreement as shown in Fig.~\ref{fig:parameters_10}B.

To summarize, segregation within a binary droplet in spite of the simplifying asymmetry, triggered by selective evaporation is observed during the drying process of a 1,2-hexanediol/water mixture droplet. The small surface tension differences cannot drive a strong enough Marangoni flow on the surface to induce a high enough convection within the droplet to obtain perfect mixing. Therefore a locally high concentration of 1,2-hexanediol accumulates near the contact line of the droplet, leading to segregation. The evolution of the vorticity field indicates different life stages of the evaporating droplet. We quantitatively compare the experimental data with a numerical simulation, showing excellent agreement. While the model perfectly predicts the water and diol concentrations in the inner center and outer layer of the droplet, respectively, note that it cannot predict the phase separation of the two liquids due to the complexity of the diol's solubility in water. Indeed, 1,2-hexanediol can mix with water at any concentration without phase separation in equilibrium due to the formation of micelles-like aggregates. However, in the dynamic system of an evaporating droplet, the continuous loss of water leads to large fluctuations through mutual attractions of micelles within the new 1,2-hexanediol phase, which eventually forms the nucleation of 1,2-hexanediol~\cite{sear2014quantitative}. From an energetic point of view, it is likely that the separated 1,2-hexanediol phase has the same, or at least a very similar, chemical potential as the mixed phase in the droplet~\cite{hyman2014liquid}. Stochastic fluctuations then lead to the phase separation. Our findings offer new perspectives to understand how surfactants act in an evaporating system, and may inspire further studies of complex dynamical aspects associated with microdroplet nucleation.

\begin{acknowledgements}

The authors thank A. Prosperetti, A. Marin, H. Reinten and M. van den Berg for the invaluable discussions and NWO and Oc\'{e} for financial support. 

\end{acknowledgements}





\end{document}